\documentclass[a4,twocolumn]{revtex4}

\usepackage{graphicx}
\usepackage{textgreek}
\usepackage{multirow}
\usepackage{amssymb,amsfonts,amsmath,mathrsfs}

\begin{document}

\title{Elastic network models for RNA: a comparative assessment with molecular dynamics and SHAPE experiments}

\author{%
Giovanni Pinamonti\,$^{1}$,
Sandro Bottaro\,$^{1}$,
Cristian Micheletti\,$^{1}$,
Giovanni Bussi \,$^{1}$%
\footnote{To whom correspondence should be addressed.
Email: bussi@sissa.it
}
}

\address{%
$^{1}$
Scuola Internazionale Superiore di Studi Avanzati, International School for Advanced Studies,
265, Via Bonomea I-34136 Trieste, Italy
}

\begin{abstract}
Elastic network models (ENMs) are valuable and efficient tools for characterizing the collective
internal dynamics of proteins
based on the knowledge of their native structures. The
increasing evidence that the biological functionality of RNAs is often linked
to their innate internal motions, poses the question of whether ENM approaches can be successfully
extended to
this class of  biomolecules.
This issue %
is tackled here by considering
various families of elastic networks of increasing complexity applied to a representative set of RNAs.
The fluctuations predicted by the alternative ENMs are stringently validated by comparison against extensive molecular dynamics simulations and SHAPE experiments. %
We find that simulations and experimental data are systematically best reproduced by either an all-atom or a three-beads-per-nucleotide representation (sugar-base-phosphate), with the latter arguably providing the best balance of accuracy and computational complexity.
\end{abstract}
\maketitle

\section{Introduction}
Characterizing the functional dynamics of RNA molecules is one of the key standing
 issues in molecular biology. The interest in this topic
is spurred
by the ongoing discovery of ever new biological roles that  RNAs can have in different contexts
(see, e.g.,~\cite{breaker2014expanding} for a recent review)
and, at the same time, by the realization that
the structure $\to$ function relationship of these molecules is often related to their internal
dynamics~\cite{mustoe2014hierarchy}.
In this respect, theoretical approaches hold much potential for
complementing experiments and provide valuable quantitative insight into the functional dynamics of
RNAs.
For instance, molecular dynamics (MD) simulations with atomistic force fields have been used to reproduce experimental
measurements and aid their interpretation
(see, e.g., Refs \cite{colizzi2012rna,chen2013high,kuhrova2013computer,yildirim2013dynamic,musiani2014molecular,pan2014ion,sponer2014molecular}).
However, it may be argued that one of the most important limitations to the systematic use of atomistic MD
simulations for characterizing the behavior of RNA is their intensive computational demand.
In fact,
most if not all current MD studies are still limited to the \textmu s timescale.

For this reason, several efforts are being spent towards developing
coarse-grained approaches capable of striking a good balance between accuracy and computational
efficiency~(see, e.g., refs~\cite{ding2008ab,jonikas2009coarse,cao2011physics,denesyuk2013coarse,xia2013rna,cragnolini2013coarse,vsulc2014nucleotide}).
In this respect, it should be noted
that coarse-grained models are valuable not only because they are amenable to extensive
numerical characterization, but precisely because their simplified formulation can offer important insight
into the main physico-chemical mechanisms that underpin the behavior and properties of a
given biomolecule.

For proteins, a successful class of such simplified models are elastic networks.
These models were originally motivated by the seminal work of Tirion~\cite{tirion1996large} who showed
that the Hessian of the potential energy of a globular protein computed from an %
 atomistic force field
could be reliably reproduced by replacing the detailed inter-atomic forces by
spring-like, harmonic  interactions.
This remarkable fact was rationalized {\em a posteriori} in
terms of the large-scale character that low-energy fluctuations have in
proteins, which makes them amenable to be captured with models that are oblivious of the
details of the potential~\cite{Hinsen:1998ee,Bahar:1997hf,atilgan2001anisotropy,micheletti2004accurate,Micheletti:2013hb}.
This observation, in turn, prompted further
development of simplified harmonic models where the structural
descriptions themselves were simplified by reducing the number of interaction centers, also termed beads.
In their simplest formulation, elastic-network models (ENM) incorporate harmonic
interactions between pairs of C$_\alpha$ beads~\cite{Hinsen:1998ee,atilgan2001anisotropy,delarue2002simplified,micheletti2004accurate} while two-beads amino acid representations, e.g.~for the main- and
side-chains~\cite{micheletti2004accurate}, can predict structural fluctuations in very good accord with atomistic MD
simulations~\cite{Fuglebakk:2013hh}.

By comparison with proteins, the development and application of elastic networks aimed at nucleic acids is still relatively unexplored.
Bahar and Jernigan first applied network models to the conformational dynamics of a transfer RNA using
a model with two beads per nucleotide~\cite{bahar1998vibrational}.
Several authors further simplified this model using  a single bead
placed on the phosphorus atom~\cite{tama2003dynamic,wang2004global,van2005comparison,wang2005comparison,fulle2008analyzing,kurkcuoglu2009collective,zimmermann2014elastic}.
More recently, Setny and Zacharias suggested that the best candidate to host a single ENM bead is the
center
of the ribose sugar in the backbone \cite{setny2013elastic}.
Other ENMs with more beads per nucleotide have also been used~\cite{delarue2002simplified,wang2004global,yang2006ognm,zimmermann2014elastic}.
Most of these studies assessed the validity of different representations by focusing
on their capability to reproduce either the structural variability observed across experimental conformers
or the Debye-Waller factors from X-ray experiments.
ENM fluctuations were also compared with accurate atomistic MD simulations,
but the comparison was either limited to short time scales \cite{van2005comparison}
or to model simple double helices~\cite{setny2013elastic}. %

Towards the goal of identifying the most suitable RNA ENM, here we
assess the performance of an extensive repertoire of ENMs which are all equally viable {\em a priori}.
These models, in fact, differ for the specific single- or multi-bead representations used for each nucleotide,  as well as for the
spatial range of the pairwise elastic interactions.
As stringent term of reference we perform \textmu s time-scale
atomistic MD simulations
on RNA molecules containing canonical A-form double helices as well as nontrivial secondary
and tertiary structures.
Additionally, we introduce a procedure to compare fluctuations with selective 2$^\prime$-hydroxyl acylation analyzed by primer extension (SHAPE) experiments~\cite{merino2005rna,weeks2011exploring}.
SHAPE reactivity is empirically known to correlate with base dynamics
and sugar pucker flexibility at the nucleotide level~\cite{mcginnis2012mechanisms} and hence is, in principle, well suited for
validating predictions of RNA internal dynamics.
Recently, Kirmizialtin {\em et al}.\ have proposed a link between fluctuations of selected torsional angles and SHAPE reactivity and
used SHAPE data as an input to improve the accuracy of force-field terms in an atomistic structure-based (Go-like)
model~\cite{kirmizialtin2015integrating}.
However, to the best of our knowledge, the present study is the first attempt of using SHAPE reactivity measurements
to assess the predictive accuracy of
three-dimensional coarse-grained models or atomistic molecular dynamics simulations.

We find that the best balance between keeping the model complexity
to a minimum and yet have an accurate description of RNAs' internal dynamics is achieved when each nucleotide is described by three beads representing
the sugar, the base, and the phosphate (SBP) groups.
 Slightly better results can be obtained using the much more computationally-demanding all-atom (AA) model.
As a matter of fact, the SBP and AA elastic network models can reproduce to a very good accuracy the principal structural fluctuations as predicted from \textmu s-long atomistic MD simulations,
both in their directions and relative amplitudes.
 Additionally, they provide a satisfactory proxy
for the nucleotide-level flexibility as captured  by experimental SHAPE data.

\section{Methods}
\subsection{RNA dataset}

\begin{figure} %
\centering
\includegraphics[scale=1]{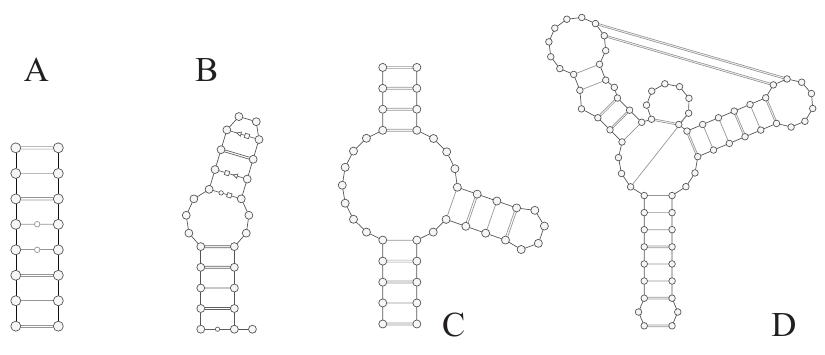}
\caption{Secondary structures of the four molecules studied:
A: eight-base-pairs duplex;
B: sarcin-ricin domain;
C: hammerhead ribozyme;
D: \emph{add} adenine riboswitch.
}
\label{fig:ss}
\end{figure}

We performed atomistic MD simulations on four different RNA molecules (Figure~\ref{fig:ss}).
These systems were chosen so as to cover a variety of
size and structural complexity and yet be amenable to extensive simulations, as detailed in Table \ref{tab:MDinfo}.

\begin{table}[htp]
\centering
\begin{tabular}{l l c c } %
\hline
 & \multirow{2}{*}{PDB code} & chain   & simulation\\
 &                           & length & time (\textmu s) \\
\hline
Duplex		& 1EKA	& 16 & $1.0$	\\
Sarcin-ricin domain 	& 1Q9A	& 25 & $0.9$	\\
Hammerhead ribozyme 	& 301D	& 41 & $0.25$ 	\\
\emph{add} Riboswitch 	& 1Y26	& 71 & $0.25$ 	\\
\emph{thiM} Riboswitch 	& 2GDI	& 78 & -  	\\
\hline
\end{tabular}
\caption{RNA dataset: details and length of MD simulations.
For the \emph{thiM} riboswitch, no MD was performed.
}
\label{tab:MDinfo}
\end{table}

The first entry is the NMR-derived model of the $^{\texttt{GAGUGCUC}}_{\texttt{CUCGUGAG}}$ RNA duplex, featuring two central G-U Wobble pairs~\cite{chen2000nuclear}.
As a second system, we considered the sarcin-ricin domain (SRD) from \emph{E.coli} 23S rRNA,
which consists of a GAGA tetraloop, a flexible region with a G-bulge and a duplex region~\cite{correll2003common}.
The U nucleobase at the 5$^\prime$ terminal was excised from the high resolution crystal structure.
We further considered two more complex molecules: the hammerhead ribozyme~\cite{scott1996capturing} and the \emph{add} adenine riboswitch~\cite{serganov2004structural}.
Both systems are composed of three stems linked by a three-way junction.
In the \emph{add} riboswitch, two hairpins are joined by a kissing loop interaction.
All these systems, except for the duplex,
were previously characterized by various computational means, including atomistic MD
simulations~\cite{vspavckova2006molecular,van2005comparison,priyakumar2010role,gong2011role,allner2013loop,di2013ligand,dipalma15bmc}.

\subsection{Molecular dynamics simulations}
All MD simulations were performed using GROMACS 4.6.7 \cite{pronk2013gromacs} with the AMBER99
force field \cite{hornak2006comparison}
including parmbsc0 \cite{perez2007refinement} and $\chi_{OL3}$ \cite{banas2010performance} corrections.
GROMACS parameters can be found at \url{http://github.com/srnas/ff}.
The trajectories were obtained in the isothermal-isobaric ensemble ($T=300$ K, $P=1$ atm) with stochastic velocity rescaling \cite{bussi2007canonical} and Berendsen barostat \cite{berendsen1984molecular}.
Long range electrostatics were treated using particle-mesh-Ewald summation \cite{darden1993particle}.
 The equations of motion were integrated with a $2$ fs time step.
All bond lengths were constrained using the LINCS algorithm \cite{hess1997lincs}.
$\textrm{Na}^+$ ions were added in the box in order to neutralize the charge,
and additional $\textrm{Cl}^-$ and $\textrm{Na}^+$ at a concentration of $0.1$ M.
AMBER-adapted parameters were used for Na$^+$ \cite{aaqvist1990ion}
and Cl$^-$ \cite{dang1995mechanism}.
The adenine ligand bound to the \emph{add} riboswitch was parametrized using the general Amber force field (gaff)~\cite{case2004amber}
and partial charges
were assigned
as discussed in reference \cite{di2013ligand}.
The analyses of the hammerhead ribozyme and of the \emph{add} riboswitch trajectories were performed after
 discarding the first $10$ ns and $5$ ns, respectively.

\subsection{Elastic Networks}

In elastic network models a simplified structural representation is
achieved by representing any monomeric unit of the biopolymer with one or
more beads.  Accordingly, the model potential energy is equivalent to the one of a set of $N$ beads connected by pairwise harmonic springs which penalize deviations of inter-bead distances from their typical, reference values.
Thus, the elastic network does not directly restrain the absolute positions of the beads but only their distances.
In the simplest formulation, the spring constant of the harmonic pairwise interaction
is set equal to a master spring constant $k$ whenever the reference distance between the two beads
is smaller than a pre-assigned interaction cutoff ($R_c$), %
 and set to zero otherwise.

The potential energy of the system can be approximated to second order as
\begin{align}
U(\delta r_{i,\mu} , \delta r_{j,\nu}) \approx {1 \over 2}  \delta r_{i,\mu} M_{i j, \mu \nu} \delta r_{j,\nu}
\label{eq:ENMpot}
\end{align}
where the $3N\times3N$ symmetric matrix, $M$, is
the Hessian of $U$, and $\delta r_{i,\mu}$ is the $\mu$ Cartesian component of the deviation of bead $i$ from its position in the reference structure.

\subsubsection{Repertoire of possible elastic networks for RNAs}

In protein contexts, the standard formulation of elastic network models is
based on the intuitive amino acid representation with primary interaction centers located on the mainchain (e.g. the C$_\alpha$ atoms)
and possibly auxiliary ones for the sidechains~\cite{micheletti2004accurate}.
By analogy with the case of proteins, one may expect that the primary ENM interaction centers could be the phosphate groups, which provide the backbone connectivity for single RNA
strands~\cite{tama2003dynamic,wang2004global,van2005comparison,wang2005comparison,fulle2008analyzing,kurkcuoglu2009collective,zimmermann2014elastic}.
Besides this possibility, we here investigated alternative representations considering all possible ENMs combinations based on the use of one or more interaction centers representing the three chemical groups of each nucleotide: the sugar, the base and phosphate (in short S, B and P, respectively).
Each group is represented by a specific atom, namely C1$^\prime$ for the sugar, C2 for the base and P for the phosphate group.
This choice is largely dictated for convenience of comparison with earlier studies~\cite{delarue2002simplified,yang2006ognm,setny2013elastic,zimmermann2014elastic}.
For each model the interaction cutoff distance, $R_c$, is varied in the $3-30$~\AA~range with $1$~\AA~increments so as to assess the dependence of the predictions on the degree of connectivity of the elastic network.

\subsubsection{Reference structure}

For each RNA dataset entry, the reference structure for ENM calculations is set equal to the
centroid structure of the associated MD trajectory. This is the conformer with the lowest average
mean square distance from
all MD-sampled structures after an optimal rigid structural alignment~\cite{kabsch1976solution}.
In the case of the \emph{add} riboswitch, the adenine ligand atoms are included in the ENM calculation.

\subsection{Comparison of ENMs and MD}

For a detailed and stringent comparison of ENM and MD we shall consider the covariance matrix,
which provides information on the structural fluctuations at equilibrium.
The MD covariance matrix entries are defined as
$C^{\text{MD}}_{i j, \mu \nu}  = \langle \delta r_{i,\mu} \delta r_{j,\nu} \rangle \;, \; \; \textrm{with} \quad
\delta r_{i,\mu} = (r_{i,\mu} - \langle r_{i,\mu}\rangle)
$
where $i$ and $j$ run over the $N$  indexed interaction centers, $\mu$ and $\nu$ run over the Cartesian components, and $\langle \rangle$ denotes the time-average over the sampled conformations after an optimal structural superposition over the reference structure.
When comparing with a coarse-grained ENM,
the structural alignment and the calculation of $C^{\text{MD}}$ are both
performed by exclusively considering the same atom types used as beads in the elastic network model.
For ENM, the covariance matrix is obtained from the pseudoinverse $\tilde{M}^{-1}$ of the interaction matrix defined in Eq.~\ref{eq:ENMpot}, as $C^{\text{ENM}}_{i j, \mu \nu} = k_B T \tilde{M}^{-1}_{i j, \mu \nu}$.
Here $k_B$ is the Boltzmann constant and $T$ is the temperature. We observe that the
$k_BT$ term is here required to allow the absolute covariance matrix to be properly
related to the spring stiffness $k$.
However, since in all the comparisons discussed below we always consider
a multiplicative term in the covariance matrix as a parameter for the fitting
procedures, the values of both 
$k_BT$ and $k$ is never used in practice.

\subsubsection*{Effective Interaction Matrix}

When comparing different ENMs one must consider only the modes related to the fluctuations of the degrees of freedom in common between the models.
To achieve this, it is necessary to separate the degrees of freedom of the beads of interest (with subscript {\em a} in the following) from the others (with subscript {\em b} in the following) and compute the effective interaction matrix of the former~\cite{zheng2005probing,Ming:2005tp,Zen:2008bt,Micheletti:2013hb}. This is accomplished by formally recasting  the interaction in the following block form
$
M=
 \left( \begin{array}{c|c}
M_a & W  \\
\hline
W^T & M_b  \\
 \end{array} \right)
$
where $M_a$ and $M_b$ are the interaction matrices of the two subsystems, while $W$ represents the interactions
between them. The effective interaction matrix governing the dynamics of subsystem $a$ alone is
\begin{align}
 M^{\textrm{eff}}_a = M_a - W M_b^{-1} W^T
\label{eq:trac}
\end{align}
For a detailed derivation of this equation see~\cite{Zen:2008bt}.
Using this effective matrix one can compute the fluctuations relative to the subsystem considered.

\subsubsection{Measures of similarity between essential spaces}

The comparison of the essential dynamical spaces of ENM and MD simulations is here carried out by considering two quantities, namely the Pearson correlation of mean square fluctuations profiles and the similarity between the eigenspaces of covariance matrices.

The mean square fluctuation (MSF) of a given center, $i$,
can be obtained in the MD simulation by time-averaging the mean square displacements.
Similarly, in ENMs they are
given by $\text{MSF}_i = \langle \delta r_i^2 \rangle = k_B T \sum_{\mu=1}^3  \tilde{M}^{-1}_{ii,\mu\mu}$.
We remark that the amplitudes of fluctuations are known to be inversely-correlated to the local density, that is the number of neighboring centers \cite{halle2002flexibility}. We also recall that the MSF profile is computed after carrying out an optimal global structural superposition of all sampled conformers. As a consequence, the MSF of any given center depends not only on the local structural fluctuations but on the global intra-molecular ones too.

The accord of two covariance matrices, $A$ and $B$, can be measured more directly by comparing their essential dynamical spaces, identified by the set of their eigenvectors $\{\mathbf{v}_A\}$ and $\{\mathbf{v}_B\}$ and eigenvalues $\{\lambda_A\}$, $\{\lambda_B\}$.
A stringent measure of this consistency is the root weighted square inner product (RWSIP)~\cite{Carnevale:2007do}
\begin{equation}
  \text{RWSIP}= \sqrt{\frac {\sum_{i,j=1}^{3N} \lambda_{A,i} \lambda_{B,j} (\mathbf{v}^i_A \cdot \mathbf{v}^j_B )^2 } {\sum_{i=1}^{3N} \lambda_{A,i} \lambda_{B,i}} }
\label{eq:rwsip}
\end{equation}
which takes on values ranging between $1$, when the two ranked dynamical spaces coincide, and $0$, when they are completely orthogonal.

The statistical significance of both the MSF correlation and the RWSIP is assessed by using two terms of reference. The first one is given by the degree of consistency of the MSF or RWSIP for first and second halves of the atomistic MD trajectories. This sets, in practice, an upper-limit for very significant correlations of the observables.
The second one is the degree of consistency of the random elastic network (RNM) of Setny {\em et al.}\cite{setny2013elastic}
with the reference MD simulations.
This is a fully-connected elastic network where where all pairs of beads interact harmonically though, for each pair, the spring constant is randomly picked from the $[0,1]$ uniform distribution.
Because this null ENM does not encode properties of the target molecule in any meaningful way, it provides a practical lower bound for significant correlations between ENMs and MD simulations.

\subsection{Comparison with SHAPE data}

To compare the fluctuations from both ENMs and MD simulations with data from SHAPE experiments
we here scrutinize several order parameters that, {\em a priori} could be viable proxies for SHAPE reactivity data, namely:
i) the variance of the distance between selected pairs of beads and ii) the variance of the angle
between selected triplets of beads.
The variance of each distance and angle as obtained from MD was compared with the SHAPE reactivity of the corresponding nucleotide
for the \emph{add} riboswitch taken from ref~\cite{rice2014rna}.%
Distances and angles were computed using PLUMED \cite{tribello2014plumed}.

In the ENM framework, the variance of the distance between two beads can be directly obtained from the covariance matrix in the linear perturbation regime as
\begin{equation}
\sigma_{d_{i j}}^2 = \sum_{\mu,\nu=1}^{3} \frac { \tilde{d}_{i j}^{\mu} \tilde{d}_{i j}^{\nu}}{\tilde{d}^2} ( C_{i i,\mu\nu} + C_{j j,\mu\nu} - C_{i j,\mu\nu} - C_{j i,\mu \nu})
\label{eq:distfluc}
\end{equation}
where $\tilde{d}_{i j}^{\mu}$ is the $\mu$ Cartesian component of the  reference distance between bead $i$ and $j$.

When comparing ENM and SHAPE we also considered the experimental data relative to the
\emph{thiM} thiamine pyrophosphate riboswitch
published in ref~\cite{rice2014rna}.
For this molecule no reference MD simulation was performed and
ENMs were computed directly on the crystal structure (PDB code: 2GDI)~\cite{serganov2006structural}.

\section{Results}

\begin{figure} %
\centering
\includegraphics[scale=1]{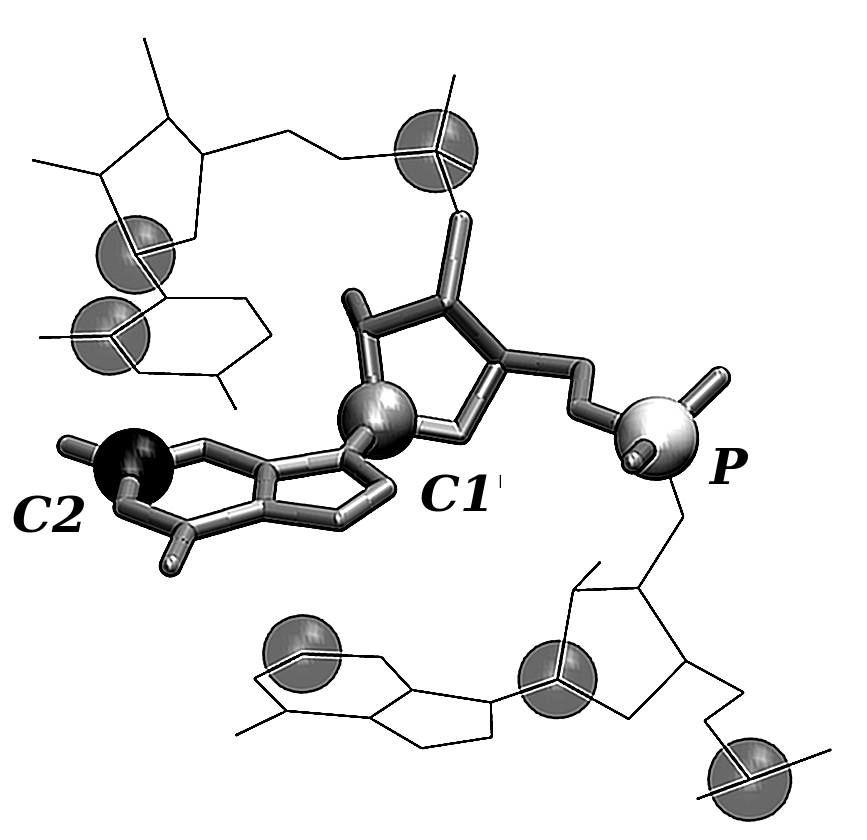}
\caption{Schematic representation of the beads used to construct the ENM. The three atoms used as beads are the C2 carbon in the base,
the C1$^\prime$ carbon in the sugar ring, and the P atom in the phosphate group, as indicated by labels.}
\label{fig:schema}
\end{figure}

For the comparative validation against MD and SHAPE data we consider eight different types of elastic networks,
as summarized in Table~\ref{table:Rc}. A subset of the considered models have been previously used in different contexts~\cite{delarue2002simplified,yang2006ognm,setny2013elastic,zimmermann2014elastic}. With the exception of the all-atom (AA) model, all other ENMs will be referred to with the one, two and three-letter acronyms corresponding to which of the phosphate (P), sugar (S), or base (B) interaction centers are used, see Fig.\ \ref{fig:schema}.
We also tested ENMs with a higher number of beads (see Fig.~SD~1 for an example).
All the considered ENMs feature a sharp-cutoff interaction scheme (as explained in the section Methods). Using a distance-dependent elastic constant yields similar results (Fig.~SD~2 for details).

\subsection{Comparison of ENMs and MD}

\begin{figure*} %
\centering
\includegraphics[scale=1.]{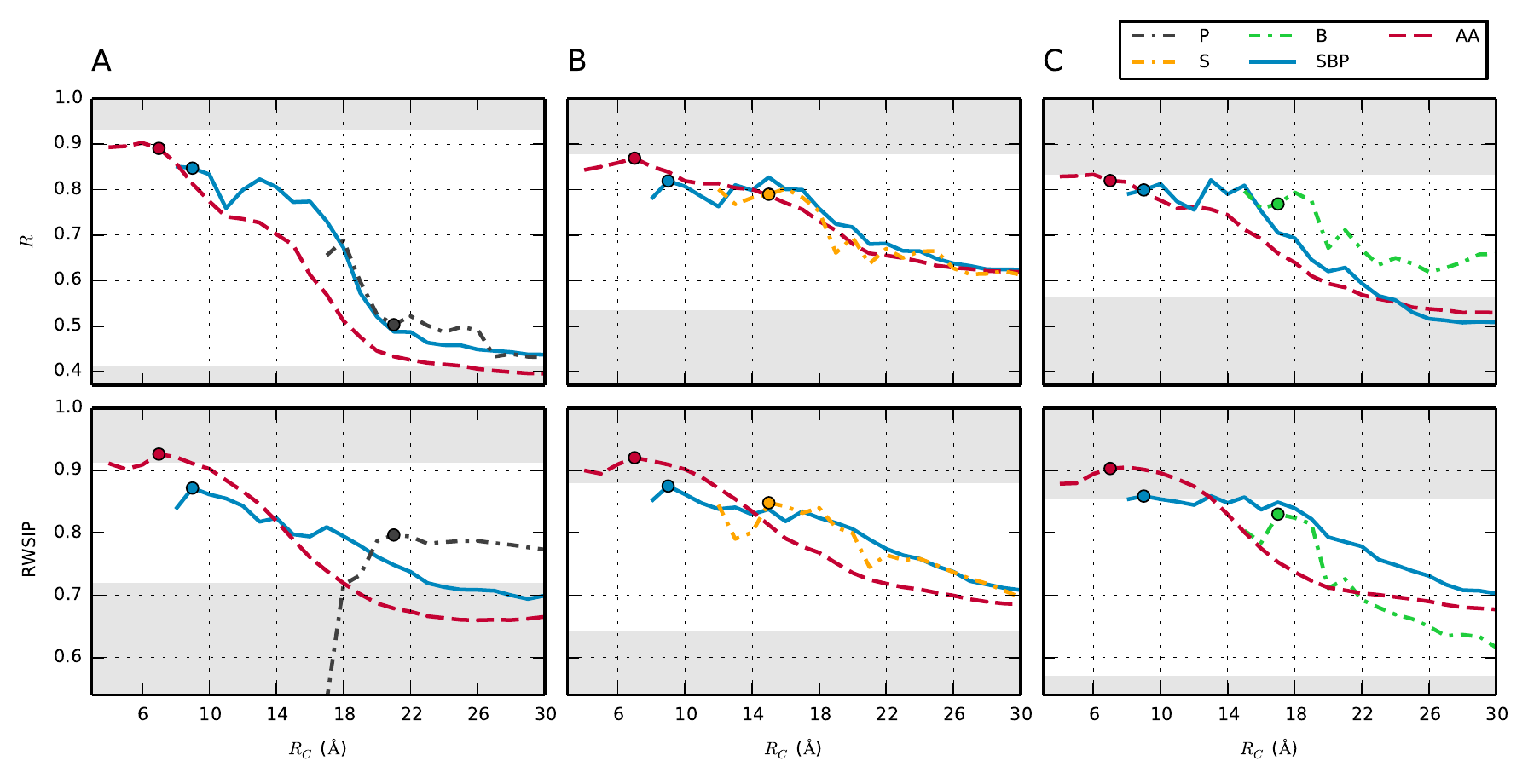}

\caption{Agreement between MD simulations and ENM for different radii of cutoff. Correlation between MSF (upper panels), and RWSIP (lower panels). Values at the optimal cutoff values are represented by circles.
  A: phosphate beads; B: sugar beads; C: nucleobase beads.
The gray regions correspond to values below the random-network model or above the MD self-agreement.}
\label{fig:enms}
\end{figure*}

The consistency of ENM and MD simulations was assessed by computing the
Pearson correlation coefficient ($R$) for the MSF profiles and the RWSIP for the essential dynamical spaces.
To keep the comparison as simple and transparent as possible,
each measure was computed separately for the S, B and P interaction centers.
For multi-center ENMs this required the calculation of the effective interaction matrix (Eq.~\ref{eq:trac}).
Using as a reference the experimental structure in place of the MD centroid introduces only minor differences in the results, see SD 3.
Each measure was then averaged over the four systems in Table \ref{fig:ss} (see SD 4 for non-averaged values).
The results, shown in Fig.~\ref{fig:enms}, are profiled as a function of the elastic network interaction cutoff distance, $R_c$.
The smallest physically-viable value for $R_c$, that is the abscissa of the left-most point of the curves,
is the  minimum value  ensuring that the ENM zero-energy modes exclusively correspond
to the six roto-translational modes.

The main feature emerging from Fig.~\ref{fig:enms} is that, across the various models,
the highest consistency with MD is attained when $R_c$ is marginally larger than its smallest physically-viable
value. It is also noted that the minimum value of $R_c$ varies significantly across the models:
for the AA model, which is the most detailed ENM, it is as low as $4$~\AA,
while for the single-bead ones it
is often larger than $10$~\AA. The MSF and RWSIP accord both decrease systematically as $R_c$ is increased
starting at the optimal value.
This fact, which to our knowledge has not been reported before, can be rationalized {\em a posteriori} by considering that upon increasing $R_c$, one endows the network with harmonic couplings among nucleotides that are too far apart to be in direct physical interaction, and this brings about a degradation in model performance.

Furthermore, it is noted that the detailed, but also computationally more onerous, AA model is consistently in better accord with MD data than any of the coarse-grained ENMs.
For this model, the degree of ENM-MD consistency is practically
as high as the internal MD consistency at the optimal value $R_c\approx7$~\AA , or even higher
in some cases.
As a general trend, we notice that the accord between MD and ENMs decreases
for coarser models (see also Fig.~SD~5 for models including two beads per nucleotide).
Importantly, the AA and SBP models perform well not only on average but for each considered structure, whereas
the performance of models with fewer interactions centers is less consistent across the repertoire of RNA molecules, see Fig.~SD~4.
For all models, considering the optimal value of $R_c$
both MSF and RWSIP accord are significantly higher than for the null
model, indicating that all the ENMs are overall capable to capture
the salient physical interactions of the system.

It is important to mention here that in the MD simulation of the duplex we observed a fraying
event at time $\approx 670$ ns (see Fig.\ SD~6), followed by a re-zipping into the native structure.
As a matter of fact, fraying events are expected at RNA termini on the \textmu s  time-scale covered by our simulations \cite{zgarbova2014base}.
In spite of the fact that these events are clearly out of the linear
perturbation regime where one would expect ENM to properly predict fluctuations,
the correlation between MD and ENM is reasonably high. By removing from the analysis the highly fluctuating terminal base pairs, the correlation is
further improved (Fig.\ SD~7).

In Table~\ref{table:Rc} we summarize all the results for the optimal cut-off radius, determined as the
radius that maximizes the RWSIP. The last column of the table reports the average number of neighbors of a bead, that is the number of other beads at distance smaller than $R_c$ from it.

\begin{table}[htp]
  \centering
  \begin{tabular}{ l  c c c c  c  c }
    \hline
    \multirow{2}{*}{ENM} &  \multirow{2}{*}{C1$^\prime$} &  \multirow{2}{*}{C2} &  \multirow{2}{*}{P} &  \multirow{2}{*}{others} &  best   & number of \\
                         &                       &                      &                     &                          & $R_c$ (\AA) & neighbors \\
    \hline
    P   &   &   & $\checkmark$ &   & $20$ & $15.3$ \\
    S	& $\checkmark$ &   &   &   & $15$ & $9.9$ \\
    B	&   & $\checkmark$ &   &   & $17$ & $14.8$ \\
    SP  & $\checkmark$ &   & $\checkmark$ &   & $19$ & $30.4$ \\
    BP	&   & $\checkmark$ & $\checkmark$ &   & $18$ & $29.9$ \\
    SB	& $\checkmark$ & $\checkmark$ &   &   & $11$ & $15.4$ \\
    SBP	& $\checkmark$ & $\checkmark$ & $\checkmark$ &   & $9$ & $12.0$ \\
    AA	& $\checkmark$ & $\checkmark$ & $\checkmark$ & $\checkmark$ & $7$ & $52.9$  \\
    \hline
  \end{tabular}
\caption{
Summary of the tested ENMs. For each model, the adopted beads are marked.
AA include all heavy atoms.
Values of the cutoff radius ($R_c$) that maximize the RWSIP
and average number of neighbors are also shown.
}
  \label{table:Rc}
\end{table}

\subsubsection{Effect of ionic strength}

One standing question for RNAs, that is relevant also for ENM development~\cite{zimmermann2014elastic}, is whether and how the internal dynamics of these biomolecules is affected by the concentration and type of counterions in solutions.
These parameters, in fact, modulate the screening of the electrostatic self-repulsion of RNA backbone and are indeed often used to artificially induce RNA unfolding.
Because current formulations of ENMs, including those considered here, do not explicitly account for
electrostatic effects, and thus intrinsically provide results that are independent of the ionic strength,
it is important to ascertain to what extent changes of ionic strength would affect the
collective internal dynamics of the considered RNAs.

To clarify this point, we carried out MD simulations at different nominal concentrations of monovalent salt Na$^+$/Cl$^-$.
The consistency of the essential dynamical spaces observed in simulations based on different salt concentrations
was measured with the RWSIP.
Only the C2, C1$^\prime$ and P atoms were considered for computing the essential dynamical spaces.

\begin{table}[htp]
  \centering
  \begin{tabular}{l c c c c}
 Molecule & $0.0$ M & $0.1$ M & $0.5$ M & $1.0$ M \\
    \hline
    Duplex &$0.938$ & $0.998$ & $0.991$ & $0.990$ \\
    SRD & $0.983$ & $0.983$ & $0.982$ & $0.993$
  \end{tabular}
  \caption{RWSIP between $100$ ns trajectories at different NaCl concentrations and the $500$ ns
trajectory at $0.1$ M.
For the duplex, only the first half of the  $1$ \textmu s trajectory was considered, thus discarding the contribution of the base fraying event (see SD 6).}
  \label{table:ion}
\end{table}

As summarized in Table~\ref{table:ion},
the essential dynamical spaces are very consistently preserved over a wide range of ionic strengths.
This finding complements a recent study of Virtanent {\em et al.} \cite{virtanen2014ionic}
where the electrostatic free energy was shown to be minimally affected by ionic strength.
In the present context, the result justifies the use of RNA elastic networks with no explicitly treatment of
the ionic strength.
It is however important to note that our test was limited to monovalent cations. The treatment of
divalent cations is known to be very challenging because of force-field limitations and sampling difficulties.

We finally notice that in our simulations with standard AMBER ions we did not
observe any ion-crystallization event~\cite{auffinger2007spontaneous}.
For maximum robustness we tested the alternative ion parameterization by Joung and Cheatham~\cite{joung2008determination}, obtaining very similar results.

\subsection{Comparison with SHAPE data}

To complement the validation of ENM against MD, we assessed their consistency with experimental data too.
To this purpose we considered data obtained from SHAPE experiments, which probe RNA structural fluctuations at the nucleotide level~\cite{mcginnis2012mechanisms}.
One standing challenge is that it is not yet settled which simple structural or dynamical observables can be used as viable proxies for the SHAPE intensities. To tackle this elusive problem, we first set out to analyze the MD simulations so as to identify the local fluctuations that best correlate with SHAPE data. Specifically, we compared our MD simulation and available SHAPE data for the \emph{add} riboswitch \cite{rice2014rna}. %
A related comparison based on B-factors profiles, which are commonly used to validate ENM predictions (albeit with known limitations \cite{Fuglebakk:2013hh}) is provided in fig.~SD~8.

\begin{figure} %
\centering
\includegraphics[width=\columnwidth]{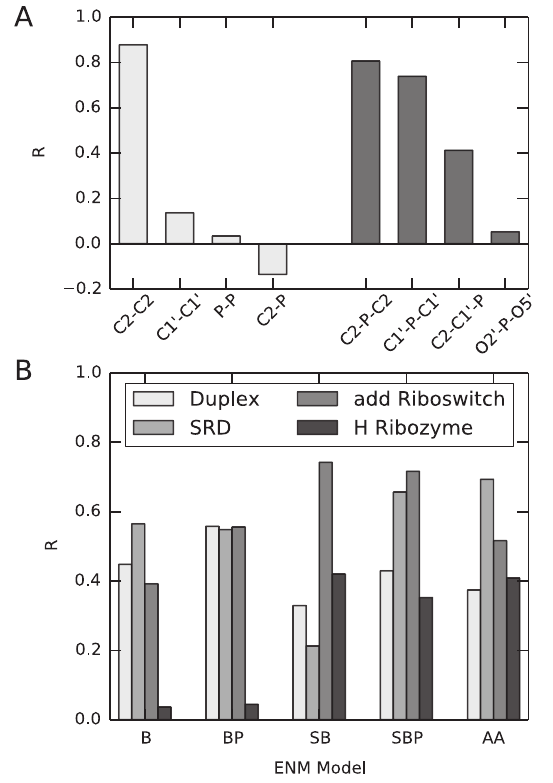}
\caption{A: Pearson correlation coefficient $R$, computed  between SHAPE reactivities and the fluctuations of different distances (light grey), and angles (dark grey), computed from the MD trajectory of the \emph{add} riboswitch.
Residue indexes are shown in Fig.~SD~10;
B:~correlation between the fluctuations of the distance of consecutive C2 atoms, from the MD simulation and from the different ENMs.}
\label{fig:MDvsENM}
\end{figure}

\begin{figure} %
\includegraphics[width=\columnwidth]{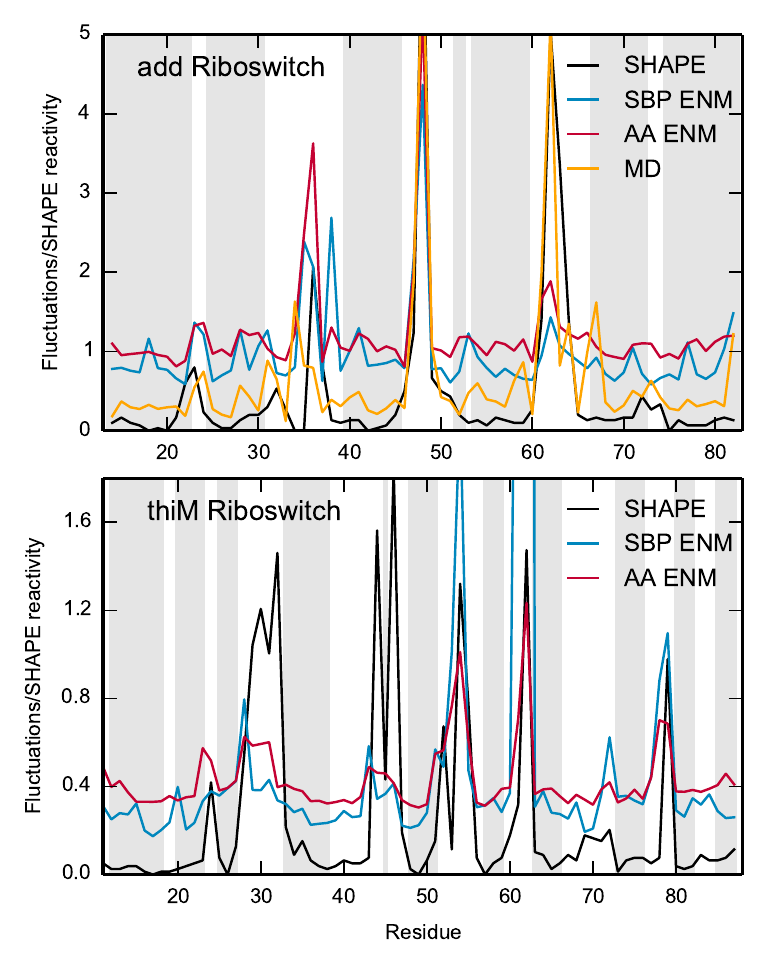}
\caption{Comparison of the flexibility of the \emph{add} riboswitch (upper panel) and the \emph{thiM} riboswitch (lower panel). The SHAPE reactivities (black) are compared with the C2-C2 fluctuations predicted by the SBP and the AA models. For the \emph{add} riboswitch, also fluctuations from MD are shown. Regions corresponding to residues forming 
Watson-Crick or wobble base
pairings are shown in gray.
}
\label{fig:ENMvsSHAPE}
\end{figure}

As it emerges from Fig.~\ref{fig:MDvsENM}A, the best correlation with experimental SHAPE reactivity
was found for the fluctuations of the distance between consecutive C2 atoms ($R=0.88$).
This is remarkable, since the  SHAPE reaction does not explicitly involve the nucleobases. %
These fluctuations are shown, as a function of the residue index, in Figure~\ref{fig:ENMvsSHAPE}.
The result can be interpreted by considering that most of the structural constraints in RNA originates from
base-base interactions, and fluctuations in base-base distance are required for backbone flexibility.
The fluctuations of the angle O2$^\prime$-P-O5$^\prime$ instead
showed a poor correlation with experimental SHAPE data ($R=0.05$).
We notice here that the value of this angle has been shown to correlate with RNA stability
related to in-line attack~\cite{soukup1999relationship},
and its fluctuations were recently used in the SHAPE-FIT approach to optimize the parameters
of a structure-based force-field
using experimental SHAPE reactivities~\cite{kirmizialtin2015integrating}.
We also observe that the fluctuations of the distance between consecutive C2 atoms could be
correlated with ribose mobility, which in turn depends on sugar pucker~\cite{altona1972conformational,altona1973conformational}.
Interestingly, C2$^\prime$-endo conformations have been shown to be overrepresented among highly reactive residues in the ribosome~\cite{mcginnis2012mechanisms}.
An histogram of C2-C2 distances for selected sugar puckers is shown in Fig.~SD~9,
indicating that C2$^\prime$-endo conformations correspond to a larger variability of the C2-C2 distance.
In conclusion, although the scope of the present SHAPE profiles comparison could be affected by the limited accuracy or precision of  both experimental and MD-generated data,  the obtained results suggest that a good structural determinant for SHAPE reactivity is arguably provided by base-base distance fluctuations.  
In Fig.~SD~10 we show this comparison using a non-parametric measure of correlation.

Based on this result, we next quantified
to which extent the ENMs are able to reproduce the profile of fluctuations of the C2-C2 distance.
This test complements the assessment made using MSF and RWSIP, which
mostly depends on the agreement of large scale motions and does not
imply a good performance in the prediction of local fluctuations.
This comparison is presented in  Figure~\ref{fig:MDvsENM}B where the ENM-MD Pearson correlation coefficients for each considered ENM
are summarized.
We remark here that the duplex (1EKA) is undergoing a base fraying, so that MD exhibits very large fluctuations
at one terminus (see Fig.\ SD~6).
The overall accord between MD and ENM is moderately good, although
significantly worse than the accord with the large scale motions presented before.
Overall, it is seen that the both the SBP model and the AA models
provide the best agreement.

In the following, we thus test whether the SBP and AA models are capable of reproducing
SHAPE reactivities directly, without the need for an expensive MD simulation to be performed.
ENM and SHAPE data were compared for two different molecules, namely
the aforementioned \emph{add} riboswitch and the \emph{thiM} riboswitch.
As we can see from Figure~\ref{fig:ENMvsSHAPE} the predictions of ENM are in qualitative agreement with the SHAPE data.
In particular, high SHAPE reactivity in the loop and junction regions correspond to highly fluctuating beads,
both for the \emph{add} and \emph{thiM} riboswitch.
We notice that this agreement goes beyond the mere identification of the residues involved in Watson-Crick or wobble
 pairings
\cite{hajdin2013accurate},
as there appear several unpaired bases with a low SHAPE reactivity.
This feature seems to be often correctly reproduced by the C2-C2 fluctuations profile.
By visual inspection, it can be seen that non-reactive, non-paired bases often engage non-Watson-Crick base pairs as well as stacking interactions, as shown in Fig.~SD~11. 
The Pearson correlation coefficients are summarized in Table~\ref{table:corrshape}. In this case too, it is found that the AA ENM  performs
 better than the SBP ENM which, nevertheless, is much less demanding computationally because of its simpler formulation.
\begin{table}[htp]
  \centering
  \begin{tabular}{c c c c}
 Molecule & SBP & AA  & MD\\
    \hline
    \emph{add} & $0.64$ & $0.76$ & $0.88$ \\
    \emph{thiM} & $0.37$ & $0.59$ &  -
  \end{tabular}
  \caption{Pearson correlation coefficients between C2-C2 fluctuations predicted by ENM/MD and SHAPE reactivities.
}
  \label{table:corrshape}
\end{table}

\section{Discussion}

The development and performance assessment of elastic networks for RNAs have so far been pursued in two main directions. On one hand, Zimmermann and Jernigan~\cite{zimmermann2014elastic} have recently shown
that the essential dynamical spaces of ENMs based on the phosphate representation of RNAs
can satisfactorily account for the structural variability observed across crystal structures homologs.
On the other hand, Setny and Zacharias~\cite{setny2013elastic} have considered ENMs where different atoms of the RNA backbone (i.e.~sugar and phosphate groups only) are alternatively used  to represent nucleotides in short RNA duplexes. Within this class of single-bead ENMs and target RNA structures, it was found that those based on the sugar-group representation yielded the structural fluctuations with the best consistency with MD simulations or NMR ensembles~\cite{setny2013elastic}.

Here, we tackle this standing challenge by searching for
the simplest and yet accurate RNA ENM. We analyze a comprehensive combinations of (i) interaction centers, or beads, for each nucleotide and (ii) spatial range of the elastic interaction. In total, we considered eight different types of ENMs, which are listed in Table~\ref{table:Rc}.
For the critical assessment of their performance, we validated the predicted structural fluctuations against data from \textmu s-long atomistic MD simulations as well as from experimental SHAPE measurements.
Finally, towards ensuring model transferability, we considered the four different types of RNA molecules listed in Table~\ref{tab:MDinfo} and represented in Fig.~\ref{fig:ss}. These systems cover a significant repertoire of different structural elements such as non-canonical base pairs, bulges, junctions and tertiary contacts and were selected with two main criteria, namely: first, they natively adopt a specific fold (i.e. have a stable tertiary structure, which is a prerequisite for ENM applicability) and, secondly, they are amenable to extensive numerical characterization with \textmu s-long MD simulations in explicit solvent. We notice that the size of the studied systems is limited only by the MD computational cost, while the ENM method is straightfowardly applicable to larger molecules, as it has been done for instance in Ref.~\cite{wang2004global}.

In the following we discuss the performance of the various models listed in Table~\ref{table:Rc} starting from those employing a single-bead nucleotide representation and then moving on to the more detailed, multi-bead ones.

Among the one-bead models the best accord with MD data is obtained for the S model, where a nucleotide is represented with the C1$^\prime$ atom of the sugar moiety. In this case, when the most appropriate elastic interaction range is used (see Table~\ref{table:Rc}), the accord of ENM and MD is significantly larger than the statistical reference (null) case, and not too much behind the accord of the first and second halves of the MD simulations.
This result is consistent with the conclusions of the aforementioned recent study of Ref.~\cite{setny2013elastic} and reinforces them from a significantly broader perspective. In fact, the present assessment is carried out for a wider range of RNA
motifs
and the search of the optimal representative atom is not limited to the RNA backbone but encompasses the base too.

In this regard, we note that the model with a single bead on the C2 atom of the base (B model)
reproduces structural fluctuations less accurately than the S model and the optimal interaction cutoff
is more dependent on the specific molecule, a fact that impairs the transferability of the model.
These shortcomings are even more evident in the P model, where a nucleotide is represented with the sole phosphorous atom. In fact, both the S and B models are better performing than the P one.
The result may be, at first, surprising because of the apparent analogy between the phosphate
representation %
in RNA and the C$_\alpha$ representation in proteins. The latter is virtually used in all single-beads ENMs for proteins.
However, one should keep in mind a fundamental distinction of backbone and side-groups roles for the structural organization and stability of these two types of biopolymers. In fact,
whereas for proteins the backbone self-interaction (e.g. hydrogen bonding) contributes significantly to the structural stability, for RNAs the analogous role is, in fact, played by the bases and not by the phosphate groups~\cite{gendron2001quantitative,bottaro2014role}. In this regard, it is interesting to recall that RNAs have, in fact,  been interpreted as adopting an ``inside-out'' organization compared to proteins~\cite{malone}. This distinction might help rationalize why the P
representation does not serve for RNA ENMs equally well as the C$_\alpha$ representation for proteins.

Moving on to two-beads models, we observe that ENMs employing beads both in the bases and in the backbone (SB, BP) perform systematically better than any single-bead model
with only a modest increase in the computational complexity. SB and BP models also outperforms the SP model.
We also stress that %
being able to reproduce the fluctuations of the bases is by itself an advantage because their functional role is of primary importance in nucleic acids and their dynamics can affect different aspects
of the behavior of RNA molecules (see, e.g., Refs.~\cite{colizzi2012rna,zgarbova2014base,bottaro2014role}).

Increasing the number of beads featured in the ENM models (see also Fig.~SD~1 for 5/6-beads model),
improves the agreement with MD, consistently with what had been observed for proteins~\cite{fiorucci2010binding}.
The best overall accuracy is indeed observed for the all-atom (AA) ENM.
We focused our attention on this model, as well as on the the SBP model, that uses one bead for each of the sugar, base and phosphate groups.
In fact, the consistency of both models with MD data is practically as high as the internal consistency of MD itself.
We also note that the optimal performance of the SBP model is attained when the interaction cutoff distance is about equal to $9$~\AA.
This is a convenient %
feature,
as this interaction range falls in the same viable interaction range of elastic networks for proteins~\cite{Fuglebakk:2013hh,micheletti2004accurate}.
Furthermore, the typical density of beads in protein ENM is very similar to the SBP model (Table \ref{table:Rc}).
In principle, this allow for the perspective of integration of proteins and RNA elastic networks to study protein/RNA complexes.

 The viability of the SBP and AA models is independently underscored by the comparison against experimental SHAPE data, which are notoriously challenging to predict.
The challenge is at least partly due to the difficulties of identifying from {\em a priori} considerations
structural or dynamical observables that correlate significantly with SHAPE data.
As a first step of the analysis we therefore considered various observables computed from atomistic MD simulations against SHAPE data,
and established that the relative fluctuations of consecutive nucleobases provide a viable proxy for SHAPE data.
Our comparative analysis showed that such fluctuations can be captured well
using the  SBP ENM, and to an even better extent with the AA ENM.
Possibly, this is a step in the direction of defining a model
able to directly correlate three-dimensional structures
with SHAPE reactivities.
Interestingly,
both the ENMs are completely independent from
the dihedral potentials and thus should not be directly affected by the
pucker conformation of the ribose. The fact that they can provide a reasonable estimate
of the backbone flexibility as measured by SHAPE reactivity suggests that
the backbone flexibility is mostly hindered by the mobility of the bases.

In conclusion, elastic network models were here compared systematically with fully atomistic molecular dynamics simulations and with SHAPE reactivities. We found that, in spite of their simplistic nature, the three-center (SBP) and all-atom (AA) elastic networks are capable of properly reproducing both MD fluctuations and chemical probing experimental data.
Of these two accurate ENMs, the three-center model (SBP), provides an ideal compromise between accuracy and computational complexity, given that retaining the full atomistic detail when modeling large structures,
such as the ribosome and other macromolecular RNA/protein complexes, can be computationally very demanding.

A module that implements the ENM for RNA discussed in this paper
has been included in the baRNAba analysis tool (\url{http://github.com/srnas/barnaba}).

\section{Supplementary Data}
Supplementary Data are available at NAR Online.

\section{Funding}
The research leading to these results has received funding from the European Research Council under the European Union's Seventh Framework Programme (FP/2007-2013) / ERC Grant Agreement n. 306662, S-RNA-S and from the Italian Ministry of Education, Projects of National Interest grant No. 2010HXAW77.

\subsubsection{Conflict of interest statement.} None declared.

\section{Acknowledgments}

We are grateful to Pavel Ban{\'a}{\v{s}},
Sim{\'o}n Poblete-Fuentes, and Martin Zacharias for reading the manuscript and
providing useful suggestions.
Francesco Di Palma is also acknowledged for providing the initial setup
and ligand parametrization
used for the \emph{add} riboswitch molecular dynamics.
Kevin Weeks is acknowledged for useful discussions.

\bibliographystyle{nar}

\end{document}